\pdfoutput=1

\documentclass[11pt]{article}

\usepackage{acl}   

\usepackage{times}
\usepackage{latexsym}

\usepackage[T1]{fontenc}

\usepackage[utf8]{inputenc}

\usepackage{microtype}

\usepackage{inconsolata}

\usepackage{graphicx}

\usepackage{kotex}  
\usepackage{amsmath}
\usepackage[table,xcdraw]{xcolor}
\usepackage{booktabs}
\usepackage{adjustbox}
\usepackage{multirow}
\usepackage{comment}
\usepackage{tcolorbox} 
\usepackage{float}
\usepackage{colortbl}
\usepackage[font=10pt,skip=4pt]{caption} 
\usepackage{microtype}
\newcommand{\madrec}{\textsc{MADRec}}

\definecolor{cBlue_1}{RGB}{141,211,199}
\definecolor{cBlue_6}{RGB}{13,76,109}
\definecolor{cBlue_7}{RGB}{16,106,130}
\definecolor{cBlue_8}{RGB}{19,136,160}
\definecolor{green}{HTML}{E0E5B6}

\title{\madrec: A Multi-Aspect Driven LLM Agent for Explainable and Adaptive Recommendation}

\author{
    {\bf Jiin Park}$^{1}$ \and {\bf Misuk Kim}$^{1,2}$\thanks{~~Corresponding author.} \\
    $^1$Department of Artificial Intelligence, Hanyang University \\
    $^2$Department of Data Science, Hanyang University\\
    \texttt{jiinpark@hanyang.ac.kr, misukkim@hanyang.ac.kr}
}

\begin{document}
\maketitle
\begin{abstract}
Recent attempts to integrate large language models (LLMs) into recommender systems have gained momentum, but most remain limited to simple text generation or static prompt-based inference, failing to capture the complexity of user preferences and real-world interactions. This study proposes the Multi-Aspect Driven LLM Agent (\madrec), an autonomous LLM-based recommender that constructs user and item profiles by unsupervised extraction of multi-aspect information from reviews and performs direct recommendation, sequential recommendation, and explanation generation. \madrec~generates structured profiles via aspect-category-based summarization and applies \textsc{Re-Ranking} to construct high-density inputs. When the ground-truth item is missing from the output, the \textsc{Self-Feedback} mechanism dynamically adjusts the inference criteria. Experiments across multiple domains show that \madrec~outperforms traditional and LLM-based baselines in both precision and explainability, with human evaluation further confirming the persuasiveness of the generated explanations.
\end{abstract}

\section{Introduction}

Recommender systems have become a core technology for enhancing user experience across various online platforms, primarily by predicting items a user is likely to prefer based on their interaction history with items \citep{10.1145/3343031.3351034, 10.1145/3451964.3451966, 9776660, xie2022decoupledinformationfusionsequential}.
Recently, more sophisticated recommendation methods have emerged by incorporating various information such as metadata, domain knowledge, and user review texts \citep{10.1016/j.knosys.2019.105058, PerezAlmaguer2021Content}.
However, existing models are often specialized for specific recommendation tasks, requiring new data collection and model training for each new task, leading to an inefficient structure \citep{yang2023palrpersonalizationawarellms}. This limitation hinders achieving generalizability and scalability required in real service environments. To address this, recent efforts have explored incorporating the strong representational power of Pretrained Language Models (PLMs) into recommender systems \cite{geng2023recommendationlanguageprocessingrlp}.
\begin{figure}[t]
  \begin{center}
    \includegraphics[width=\linewidth]{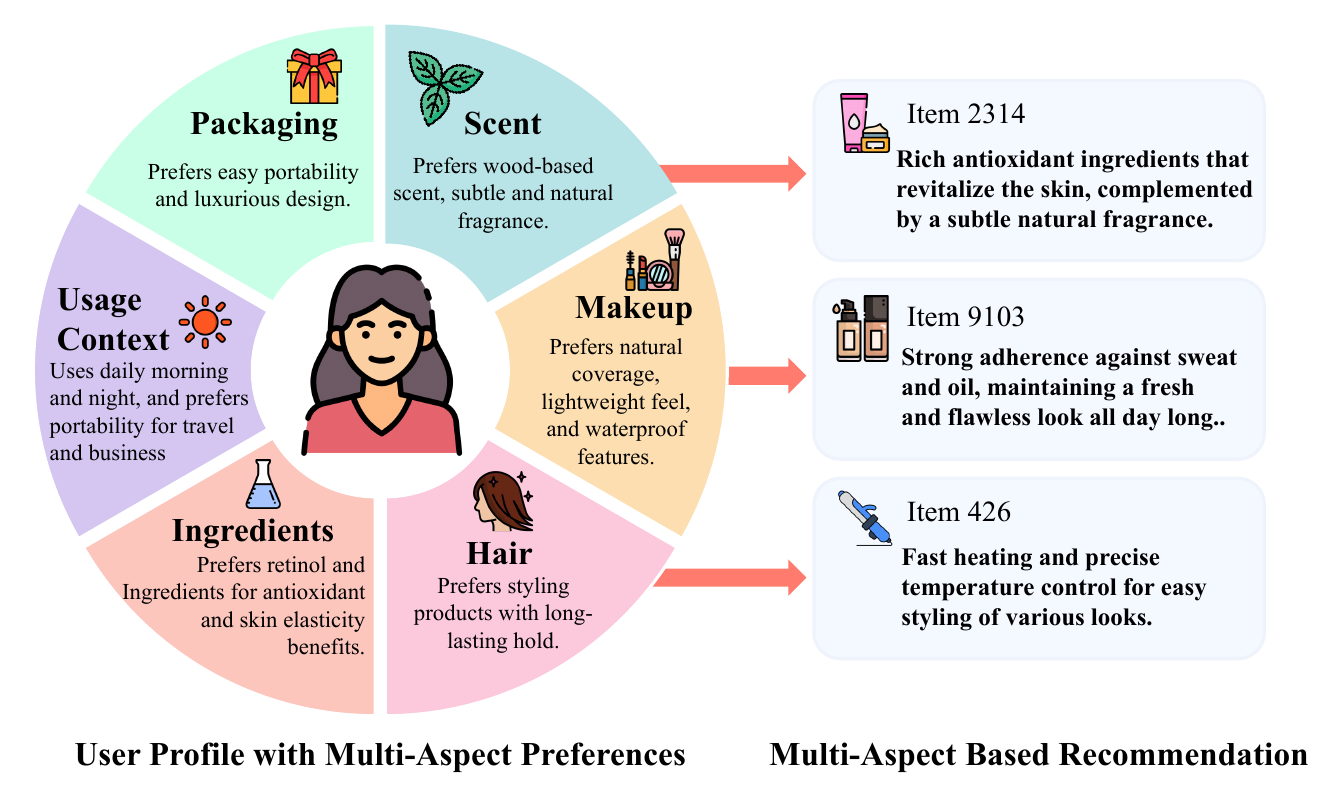}
  \end{center}
    \caption {Multi-aspect user profiles and explainable recommendations grounded in aspect-based reasoning.}
  \label{fig:introfig}
\end{figure}
In particular, Large Language Models (LLMs) such as \texttt{GPT-3}~\citep{brown2020languagemodelsfewshotlearners}, \texttt{GPT-4}~\citep{openai2024gpt4technicalreport}, and \texttt{LLaMA}~\citep{touvron2023llamaopenefficientfoundation} have significantly improved the ability to understand sentence context and reason about relationships between words and concepts through large-scale text training. These capabilities are also meaningfully applicable to recommender systems.

However, most existing research utilizing LLMs has been limited to text response generation, and high-level use cases involving tool integration, external knowledge reference, and user feedback have not been sufficiently explored~\citep{geng2023recommendationlanguageprocessingrlp}.
Moreover, users expect systems that go beyond simple item recommendations to provide explainable personalized recommendations reflecting the detailed preferences of individual users, along with persuasive explanations.
For instance, user reviews often contain information across various aspects such as texture, effectiveness, and usability in natural language expressions like \textit{\textquotedblleft It applies smoothly and has excellent pigmentation, great for dry skin\textquotedblright}~\citep{tang-etal-2024-aspect}.
Such information can serve as key clues for inferring user preferences, as well as effectively conveying the reasons behind recommendations~\citep{park2025scalableunsupervisedframeworkmultiaspect}.
However, traditional collaborative filtering and content-based approaches struggle to structure and interpret such unstructured and multidimensional text data.

In this study, we propose \madrec~(Multi-Aspect Driven LLM Agent), a framework that integrates multi-aspect-based unsupervised learning techniques with an LLM agent architecture to support a scalable, multi-domain recommendation system using LLMs (see Figure~\ref{fig:introfig}). First, aspect terms and categories are extracted from reviews using the Aspect Extraction Module. Then, reviews labeled with the same category are clustered, and category-specific summary sentences are generated using the Aspect Summary Module to construct user and item profiles.
These user and item profiles are then re-ranked through the \textsc{Re-Ranking} tool, and the top-ranked candidate items are provided as input to the LLM to generate recommendation results and explanations.
A \textsc{Self-Feedback} mechanism is applied based on recommendation results to further enhance model performance. To validate the effectiveness of the proposed framework, we conducted experiments using real review data from three domains collected from Amazon.
We conducted quantitative evaluations of our framework across three key tasks—direct recommendation, sequential recommendation, and explanation generation.
Additionally, we compared its performance against traditional recommendation models and recent LLM-based baselines in each task, demonstrating that our framework yields competitive results not only in terms of accuracy but also in explainability and user-personalized reasoning.

\noindent The main contributions of our work are:
\begin{itemize}
\item \textbf{Proposal of an unsupervised multi-aspect profile generation method}: We extract meaningful multidimensional information from unlabeled review texts and automatically generate user and item profiles, laying the foundation for explainable personalized recommendations.
\item \textbf{Introduction of aspect-based \textsc{Re-Ranking} strategy}: We design a \textsc{Re-Ranking} tool that utilizes profile information generated by the \textsc{Aspect Summary Tool} to evaluate the importance of candidate recommendation items and reorder them so that key items appear at the top of the LLM input.
\item \textbf{Implementation of an LLM-based explainable personalized agent architecture}: We construct an active agent architecture integrating reasoning, memory, tools, \textsc{Self-Feedback}, and \textsc{Re-Ranking}, enabling flexible execution of various recommendation tasks within a single framework.
\end{itemize}

\begin{figure*}[t]
  \centering
  \makebox[\textwidth][c]{%
    \includegraphics[width=1.08\linewidth]{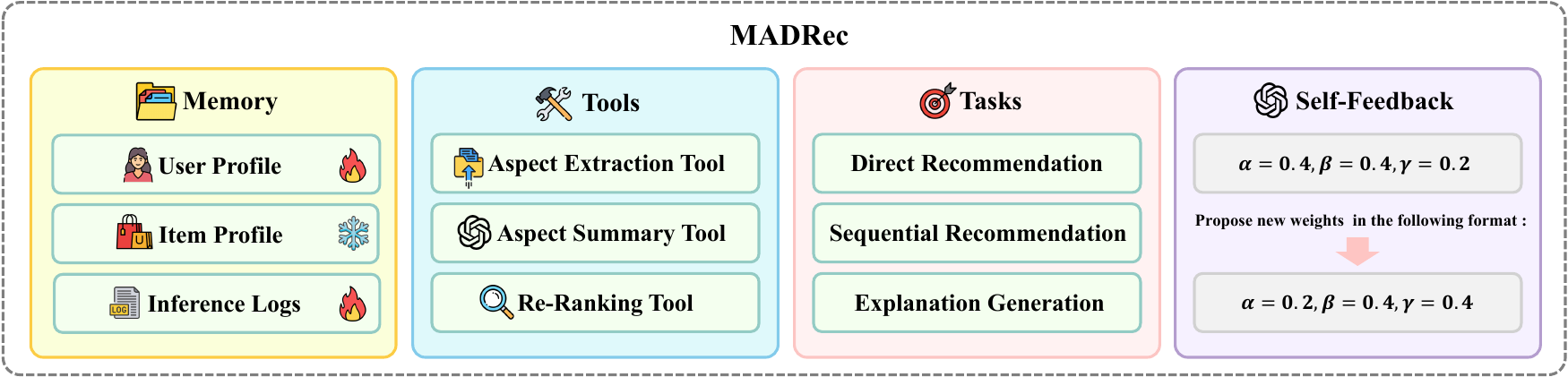}%
  }
  \caption{The structure of the \madrec~framework. The system consists of \textsc{Memory}, \textsc{Tools}, \textsc{Tasks}, and \textsc{Self-Feedback}.}
  \label{fig:1}
\end{figure*}
\section{Related Work}

\textbf{LLM-based Recommendation System}~~LLMs, leveraging their linguistic expressiveness and pretrained knowledge, are capable of understanding user preferences at the natural language level, and research efforts have increasingly aimed to integrate them into recommender systems \citep{zhang2021language, cui2022m6recgenerativepretrainedlanguage, 10.1145/3523227.3546767}. Early approaches proposed reformatting user–item interactions or metadata into sentence form, allowing recommendation tasks to be handled within a text-to-text paradigm \citep{10.1145/3523227.3546767}.
Subsequent methods modeled item attributes and user sequences as sentence-level inputs to Transformer-based architectures \citep{10.1145/3580305.3599519}. In studies where LLMs are used directly as recommenders, their performance has generally been found to be limited compared to traditional recommendation models \citep{liu2023chatgptgoodrecommenderpreliminary}, prompting follow-up work on evaluating their ability to understand personalization and on applying fine-tuning strategies \citep{kang2023llmsunderstanduserpreferences}.
Other efforts have explored prompt structures to enhance interactivity and explainability \citep{gao2023chatrecinteractiveexplainablellmsaugmented}, as well as zero-shot ranking approaches \citep{wang2023zeroshotnextitemrecommendationusing}. Fine-tuning large-scale models for personalized recommendation based on natural language user histories has also shown competitive performance \citep{yang2023palrpersonalizationawarellms}.\\
\textbf{LLM-based Agents in Recommendation Systems}~~Recent research has actively explored extending LLMs into autonomous problem-solving agents. ReAct alternates between generating thoughts and external actions to establish a sophisticated problem-solving flow \citep{yao2023reactsynergizingreasoningacting}, while Toolformer proposes a structure in which the model autonomously determines when to invoke external tools \citep{schick2023toolformerlanguagemodelsteach}. AutoGPT and BabyAGI aim to autonomously decompose high-level goals into sub-tasks, and LangChain has been utilized as a framework for implementing agent workflows \citep{AutoGPT2023, BabyAGI2023, LangChain2023}.
In the context of recommender systems, TallRec improves efficiency through domain-specific prompt tuning \citep{10.1145/3604915.3608857}, while other studies have demonstrated the potential of zero-shot ranking \citep{10.1007/978-3-031-56060-6_24} and interactive recommendation structures \citep{gao2023chatrecinteractiveexplainablellmsaugmented}.

While prior studies have largely focused on limited functionalities or static workflows, this work proposes an active agent architecture that integrates LLM reasoning capabilities, external tool usage, and a \textsc{Self-Feedback} mechanism. This enables seamless execution of multi-aspect-based user preference inference, candidate \textsc{Re-Ranking}, and explanation generation within a unified framework.

\section{\madrec~Framework}
\label{sec:madrec}
The framework proposed in this paper, which combines multi-aspect-based unsupervised learning with an LLM-based agent architecture, is illustrated in Figure~\ref{fig:1}.

\subsection{\textsc{Memory}}
\label{sec:memory}
\textsc{Memory} is a core module that stores and provides multidimensional information about users and items, allowing LLMs to reference them during recommendation tasks. The user profile is dynamically generated at each recommendation point based on reviews and purchase history, using the \textsc{Aspect Extraction Tool} and the \textsc{Aspect Summary Tool}, and is subsequently updated in \textsc{Memory}. In contrast, item profiles are constructed in advance using the same tools and are statically stored in \textsc{Memory}, simulating a real-world service environment. Detailed descriptions of these tools are provided in Section~\ref{sec:tools}.
The user and item profiles stored in \textsc{Memory} consist of summary sentences organized by aspect category. Based on these profiles stored in \textsc{Memory}, the LLM evaluates candidate items and generates recommendation explanations. Furthermore, the inference results output by the LLM during the recommendation task, as well as the weight adjustments and re-recommendation history performed in the \textsc{Self-Feedback} phase, are also logged in \textsc{Memory}. This structure enables flexible adaptation to evolving user preferences, provides essential information for LLM reasoning in a structured manner, and facilitates record-based improvement strategies for enhancing future recommendation performance.

\subsection{Tools}
\label{sec:tools}
\textbf{\textsc{Aspect Extraction Tool}}~~This is an unsupervised module that automatically extracts key aspect categories and terms from review texts. In this study, to ensure functionality without predefined labels or domain-specific formats, we apply an unsupervised clustering model to review word embeddings to group semantically similar terms, which are then used as initial candidates for aspect categories. Subsequently, multi-head attention and max-margin loss are applied to refine contextual understanding, and finally, interpretable aspect categories are assigned to each cluster by combining domain knowledge-based rules with a \texttt{GPT}-based language model. This tool is implemented with reference to the \texttt{MUSCAD} framework proposed by \citet{park2025scalableunsupervisedframeworkmultiaspect}, and is designed for extensibility across various domains. For example, in the Beauty domain, words such as \textit{evening, morning, night, daily} are grouped into the Usage Context category; \textit{aging, elasticity, reduce, dryness} into the Improvement category; and \textit{tropical, fruity, musk, sandalwood} into the Scent category. The extracted categories and terms serve as the foundational basis for constructing user and item profiles. Examples of the extracted aspect categories and terms are presented in Appendix~\ref{sec:aspect_term_and_category}, Tables~\ref{tbl:beauty_aspect_term_category},~\ref{tbl:sports_aspect_term_category}, and~\ref{tbl:toys_aspect_term_category}.\par
\noindent \textbf{\textsc{Aspect Summary Tool}}~~This tool utilizes the aspect categories and terms extracted by the \textsc{Aspect Extraction Tool} to label each review sentence with the corresponding aspect category~(Table~\ref{tbl:aspect_category_review} in Appendix~\ref{sec:aspect_term_and_category}). It then groups sentences belonging to the same category and summarizes them using an LLM on a per-category basis (Figure~\ref{fig:aspect summary}). The resulting summary sentences are stored in \textsc{Memory} as part of the user and item profiles. These summaries are subsequently included in the LLM input prompts and serve as key conditions for performing various recommendation tasks. For instance, a multi-aspect summary for a single product may appear in the form shown in Figure~\ref{fig:multi-aspect summary}.
\begin{tcolorbox}
    [title={User Profile Example},
    colback = cBlue_1!10, colframe = cBlue_7, coltitle=white, fonttitle=\bfseries\footnotesize,
    center title, fontupper=\footnotesize, fontlower=\footnotesize]
    \textcolor{teal}{Satisfaction}: Values quality, durability, and variety in nail products. \textcolor{teal}{Usage Context}: Prefers long-lasting products suitable for frequent nail changes. \textcolor{teal}{Beauty Tools}: User values durability and effectiveness for nail care products. \textcolor{teal}{Makeup}: Prefers long-lasting products with daily maintenance for durability. \textcolor{teal}{Quantity}: Prefers sets with a mix of liked and lesser plates. \textcolor{teal}{Packaging}: User values attractive and quality packaging for plates.
\end{tcolorbox}
\begin{tcolorbox}
    [title={Item Profile Example},
    colback = cBlue_1!10, colframe = cBlue_7, coltitle=white, fonttitle=\bfseries\footnotesize,
    center title, fontupper=\footnotesize, fontlower=\footnotesize]
    \textcolor{teal}{Satisfaction}: Light, soft scent loved for daily wear, despite not being show-stopping. \textcolor{teal}{Usage Context}: Customers appreciate the light and charming scent for daily wear, despite its subtle nature. \textcolor{teal}{Scent}: Delicate and charming scent, not overpowering but pleasant for daily wear. \textcolor{teal}{Purchase}: Customers repeatedly buy the fresh, dainty scent for its charm and travel-friendly packaging.
\end{tcolorbox}

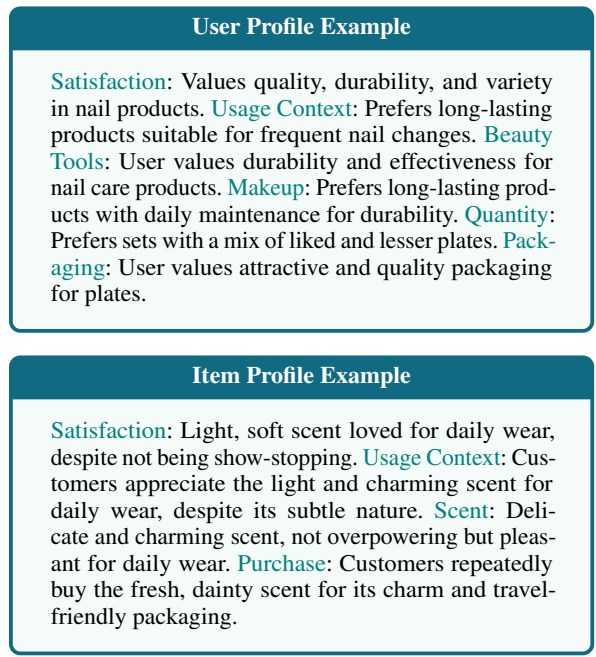
\captionof{figure}{Examples of user and item profiles constructed with our aspect-based framework. The texts highlighted in \textcolor{teal}{teal} indicate aspect categories, and the following sentences are the summary statements generated for each category.}
\label{fig:multi-aspect summary}

\textbf{\textsc{Re-Ranking} Tool}~~This tool quantifies the relevance between users and items to select the final candidate items that will be used as input for the LLM. It computes scores for items in the initial candidate pool and selects the top-$k$ items, thereby forming an input with high information density, which plays a crucial role in improving the inference quality of the LLM. This design reflects the finding that not only the inclusion but also the position of information within the input can significantly affect the accuracy of LLM outputs when processing long contexts~\citep{liu2023lostmiddlelanguagemodels}. Accordingly, the tool places high-scoring core items at the beginning of the LLM input to support more precise reasoning within the model’s limited context window.
The final score $S_i$ for a candidate item $i$ is defined as follows:
\begin{equation*}
S_i = \alpha \cdot \mathrm{Sim}(u, i) + \beta \cdot \mathrm{Sim}(C(u), C(i)) + \gamma \cdot \mathrm{Pop}(i)
\label{eq1}
\end{equation*}
\hspace*{1em}Here, $\mathrm{Sim}(u, i)$ denotes the cosine similarity between the user profile and the item profile, $\mathrm{Sim}(C(u), C(i))$ represents the similarity between the set of aspect categories associated with the user's past purchases $C(u)$ and those of the candidate item $C(i)$, and $\mathrm{Pop}(i)$ is a relative popularity indicator calculated based on the number of reviews for item $i$.\\
\hspace*{1em}In this study, we leverage multi-aspect-based user and item profiles—summarized at the aspect category level—for the \textsc{Re-Ranking} computation, enabling a finer-grained reflection of user preferences compared to simple keyword matching or frequency-based ranking. In other words, the multidimensional characteristics extracted from reviews are actively incorporated into the scoring process, effectively capturing subtle differences in individual user preferences.
\begin{figure*}[t]
  \centering
  \makebox[\textwidth][c]{%
    \includegraphics[width=1.1\linewidth]{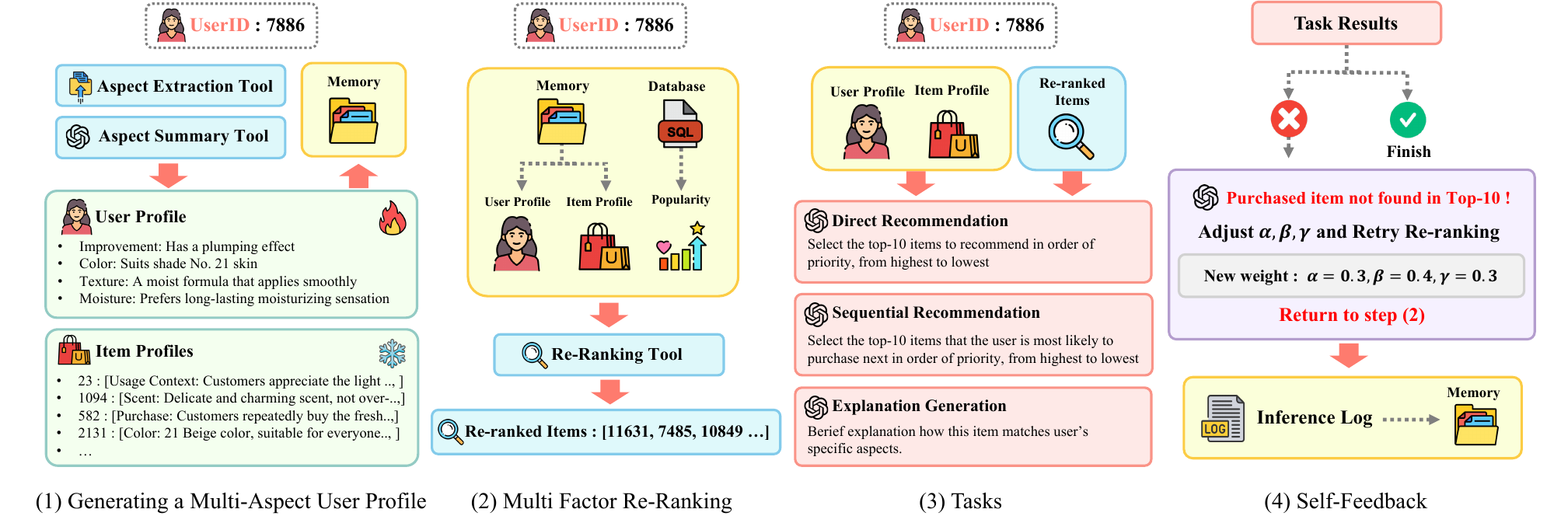}
  }
  \caption{Overall pipeline of the proposed~\madrec~framework. The system proceeds in four stages: (1) Generating a Multi-Aspect User Profile, (2) Multi Factor Re-Ranking, (3) Tasks, and (4) Self-Feedback.}
  \label{fig:pipeline}
\end{figure*}
\subsection{Tasks}
\label{sec:tasks}
The~\madrec ~framework performs three main recommendation tasks centered around the LLM: direct recommendation, sequential recommendation, and explanation generation.

\textbf{Direct Recommendation}~~This task directly recommends the most suitable items at the current point in time based on the user profile. The prompt includes the user profile and a refined list of candidate items, and the LLM selects the recommended items in order of priority and responds in natural language.

\textbf{Sequential Recommendation}~~This task predicts the items that the user is most likely to prefer next, based on their sequential purchase history. The input prompt contains the most recent five past items sorted in chronological order, the user profile, and the refined candidate item list. Based on this information, the LLM generates top recommended items.

\textbf{Explanation Generation}~~For each recommended item, the LLM generates a natural language sentence explaining why the item is suitable for the user, organized by aspect category. The LLM receives the user profile and the multi-aspect summary profile of each recommended item as input.

\hspace*{1em}Each task is formulated as a prompt that includes input information such as the user and item profiles, candidate item list, and past interactions, which is then fed into the LLM. The LLM performs reasoning in a step-by-step \texttt{Chain-of-Thought~(CoT)} manner and generates responses in natural language. The prompts used for all recommendation tasks are illustrated in Figure~\ref{fig:prompt} in Appendix~\ref{sec:Prompt}.

\subsection{\textsc{Self-Feedback}}
\label{sec:sf}
If the user’s actual purchase item is not included in the recommendation results, the \textsc{Self-Feedback} mechanism is activated. This simulates user behaviors such as re-searching or adjusting filters to find the desired product. Specifically, when the correct item is not included in the recommendation output, the weight coefficients $\alpha$, $\beta$, and $\gamma$ in the \textsc{Self-Feedback} formula are adjusted to dynamically revise the recommendation criteria, and the LLM is prompted to re-rank and recommend based on a new set of candidates. This structure enables the LLM to reflect on and refine its initial reasoning, and each step is logged and stored in \textsc{Memory}, where it can be used for future recommendations and agent decision-making. The \textsc{Self-Feedback} prompt is shown in Figure~\ref{fig:Feedback prompt} in Appendix~\ref{sec:Prompt}.

\section{\madrec-Based Recommendation Pipeline}
Based on the components described in Section~\ref{sec:madrec}, the \madrec-based recommendation pipeline consists of the following four steps, as illustrated in Figure~\ref{fig:pipeline}:\\
\textbf{Step 1. Generating a Multi-Aspect User Profile}: As shown in Figure~\ref{fig:pipeline} (a), when a recommendation request is received, the user's reviews are processed through the \textsc{Aspect Extraction Tool} and \textsc{Aspect Summary Tool} to dynamically generate a user profile organized by aspect categories. Item profiles are pre-generated in the same way and stored in \textsc{Memory}.\\
\textbf{Step 2. Multi-Factor \textsc{Re-Ranking}}: The \textsc{Re-Ranking Tool} computes a score for each item by combining the similarity between user and item profiles, category overlap, and popularity, and selects the top 30 candidate items (See Figure~\ref{fig:pipeline} (b)).\\
\textbf{Step 3. Tasks}: As illustrated in Figure~\ref{fig:pipeline} (c), the selected candidate items are used as input to the LLM, and three tasks are performed: direct recommendation, sequential recommendation, and explanation generation (see Section~\ref{sec:tasks}).\\
\textbf{Step 4. \textsc{Self-Feedback}}: If the actual purchased item is not included in the recommendation results, the \textsc{Self-Feedback} module is triggered, as shown in Figure~\ref{fig:pipeline} (d), to adjust the \textsc{Re-Ranking} weights and repeat the recommendation task.

\section{Experimental Setup}
\subsection{Datasets}
This study conducts evaluations using three real-world datasets with varying domains and levels of data sparsity. The data were collected from Amazon.com~\footnote{https://nijianmo.github.io/amazon/}, containing user reviews and ratings across a wide range of product categories. Among them, three categories—Beauty, Sports, and Toys—were selected for the experiments. After preprocessing, the statistics of each dataset are summarized in Table~\ref{tab:1}.
\begin{table}
\footnotesize
    \centering
    \resizebox{0.9\columnwidth}{!}{
        \begin{tabular}{l| ccc}
        \toprule
        Statistics & Beauty & Sports  & Toys\\
        \midrule
       \# Users & 22,363 & 25,598 & 19,412 \\
       \# Items & 12,101 & 18,357 & 11,924 \\
       \# Actions/User & 8.9 & 8.3 & 8.1     \\
       \# Actions/Item  & 16.4 & 16.1 & 14.1    \\
       \# Actions & 198,502 & 296,337 & 167,597\\
       Sparsity & 99.93\% & 99.95\% & 99.93\%\\
       \bottomrule
    \end{tabular}}
    \caption{Statistics of the datasets after preprocessing. \#Actions/User and \#Actions/Item denote the average number of interactions per user and item, respectively. Sparsity indicates the proportion of missing entries in the user-item matrix}
    \label{tab:1}
\end{table}

\subsection{Evaluation Metrics}
To quantitatively evaluate the performance of the proposed system, this study adopts a leave-one-out strategy, where one item is repeatedly excluded from each user’s interaction sequence and set as the prediction target. This approach assesses how accurately the model can predict the excluded item.
For the evaluation of direct and sequential recommendation tasks, we use HR@n (Hit Ratio) and NDCG@n (Normalized Discounted Cumulative Gain) as performance metrics, with $n$ set to 5 and 10 to account for both the hit rate and the ranking of top recommendations.
For the explanation generation task, we employ n-gram-based automatic evaluation metrics such as BLEU-n and ROUGE-n to assess the quality of the generated natural language explanations. Additionally, we use the pretrained language model-based BERT-Score to provide a more fine-grained assessment of semantic similarity.

\subsection{Baselines}
To compare the performance of the proposed model, we follow the experimental settings of \citet{10.1145/3523227.3546767, 10.1145/3340531.3411954, liu2023chatgptgoodrecommenderpreliminary} and select the following representative baseline models.\\
\hspace*{1em}For the direct recommendation task, we use \texttt{ENMF}~\citep{10.1145/3331184.3331192}, \texttt{SimpleX}~\citep{10.1145/3459637.3482297}, \texttt{P5}~\citep{10.1145/3523227.3546767}, and \texttt{ChatGPT}~\citep{liu2023chatgptgoodrecommenderpreliminary} as baselines.
For the sequential recommendation task, we include \texttt{P5}, \texttt{ChatGPT}, \texttt{S\textsuperscript{3}-Rec}~\citep{10.1145/3340531.3411954}, and \texttt{SAS-Rec}~\citep{kang2018selfattentivesequentialrecommendation}.
For the explanation generation task, we compare with \texttt{P5} and \texttt{ChatGPT}.

\hspace*{1em}Our framework uses \texttt{GPT-4.1-nano}~\citep{schulman2022chatgpt} as the core language model, and to efficiently reference domain-specific information, the entire review dataset is stored in a MySQL database. This database consists of tables that include product metadata, user interaction histories, and profile information pre-generated by the tools. Detailed descriptions of each baseline model can be found in Appendix~\ref{sec:baseline_details}.

\begin{table*}[!htbp]
\footnotesize
\centering
\label{tab:direct_beauty_sports_toys}
\resizebox{1\textwidth}{!}{
\begin{tabular}{lcccccccccccc}
\toprule
\multirow{2}{*}{Methods} & \multicolumn{4}{c}{\textbf{Beauty}} & \multicolumn{4}{c}{\textbf{Sports}} & \multicolumn{4}{c}{\textbf{Toys}} \\
\cmidrule(lr){2-5} \cmidrule(lr){6-9} \cmidrule(lr){10-13}
& HR@5 & NDCG@5 & HR@10 & NDCG@10 & HR@5 & NDCG@5 & HR@10 & NDCG@10 & HR@5 & NDCG@5 & HR@10 & NDCG@10 \\
\midrule
ENMF               & 0.020 & 0.016 & 0.050 & 0.025 & 0.096 & 0.062 & 0.144 & 0.078 & 0.066 & 0.042 & 0.128 & 0.062 \\
P5                 & 0.090 & 0.053 & 0.166 & 0.079 & 0.100 & 0.066 & 0.170 & 0.079 & 0.110 & 0.071 & 0.174 & 0.092 \\
SimpleX            & 0.040 & 0.017 & 0.082 & 0.026 & 0.034 & 0.013 & 0.054 & 0.018 & 0.050 & 0.029 & 0.086 & 0.036 \\
ChatGPT            & 0.044 & 0.029 & 0.078 & 0.040 & 0.043 & 0.082 & 0.022 & 0.035 & 0.045 & 0.025 & 0.076 & 0.035 \\
RR + SF (ours)            & \textbf{0.252} & \textbf{0.152} & \textbf{0.364} & \textbf{0.188} & \textbf{0.188} & \textbf{0.117} & \textbf{0.310} & \textbf{0.156} & \textbf{0.200} & \textbf{0.131} & \textbf{0.334} & \textbf{0.174} \\ \midrule
RR + No-SF         & \underline{0.218} & \underline{0.133} & \underline{0.296} & \underline{0.158} & \underline{0.162} & \underline{0.103} & \underline{0.264} & \underline{0.132} & \underline{0.174} & \underline{0.114} & \underline{0.260} & \underline{0.142} \\
No-RR + SF         & 0.132 & 0.090 & 0.246 & 0.126 & 0.150 & 0.098 & 0.258 & 0.132 & 0.106 & 0.070 & 0.214 & 0.104 \\
No-RR + No-SF      & 0.110 & 0.074 & 0.186 & 0.099 & 0.108 & 0.072 & 0.180 & 0.095 & 0.100 & 0.066 & 0.152 & 0.083 \\
\bottomrule
\end{tabular}
}
\caption{Performance comparison direct recommendation on Beauty, Sports, and Toys domains. Bold indicates the best score, underline the second-best. RR and SF denote \textsc{Re-Ranking} and \textsc{Self-Feedback}.}
\label{tab:direct_eval}
\end{table*}

\begin{table*}[!htbp]
\footnotesize
\centering
\resizebox{1\textwidth}{!}{
\begin{tabular}{lcccccccccccc}
\toprule
\multirow{2}{*}{Methods} & \multicolumn{4}{c}{\textbf{Beauty}} & \multicolumn{4}{c}{\textbf{Sports}} & \multicolumn{4}{c}{\textbf{Toys}} \\
\cmidrule(lr){2-5} \cmidrule(lr){6-9} \cmidrule(lr){10-13}
& HR@5 & NDCG@5 & HR@10 & NDCG@10 & HR@5 & NDCG@5 & HR@10 & NDCG@10 & HR@5 & NDCG@5 & HR@10 & NDCG@10 \\
\midrule
P5                 & 0.046 & 0.029 & 0.048 & 0.030 & 0.072 & 0.042 & 0.116 & 0.056 & 0.066 & 0.041 & 0.110 & 0.055 \\
S\textsuperscript{3}-Rec             & 0.056 & 0.034 & 0.106 & 0.049 & 0.046 & 0.025 & 0.104 & 0.043 & 0.046 & 0.027 & 0.088 & 0.040 \\
SAS-Rec            & 0.070 & 0.048 & 0.135 & 0.069 & 0.103 & 0.058 & 0.169 & 0.099 & 0.090 & 0.054 & 0.128 & 0.081 \\
ChatGPT            & 0.018 & 0.012 & 0.046 & 0.023 & 0.022 & 0.019 & 0.032 & 0.026 & 0.029 & 0.014 & 0.038 & 0.018 \\
RR + SF (ours)        & \textbf{0.234} & \textbf{0.155} & \textbf{0.362} & \textbf{0.196} & \textbf{0.230} & \textbf{0.142} & \textbf{0.368} & \textbf{0.186} & \textbf{0.202} & \textbf{0.136} & \textbf{0.336} & \textbf{0.178} \\ \midrule
RR + No-SF         & \underline{0.206} & \underline{0.142} & \underline{0.312} & \underline{0.177} & \underline{0.180} & \underline{0.115} & \underline{0.268} & \underline{0.142} & \underline{0.178} & \underline{0.120} & \underline{0.278} & \underline{0.152} \\
No-RR + SF         & 0.136 & 0.086 & 0.246 & 0.121 & 0.118 & 0.073 & 0.206 & 0.101 & 0.128 & 0.089 & 0.200 & 0.112 \\
No-RR + No-SF      & 0.104 & 0.068 & 0.188 & 0.095 & 0.104 & 0.072 & 0.140 & 0.083 & 0.104 & 0.069 & 0.144 & 0.082 \\
\bottomrule
\end{tabular}
}
\caption{Performance comparison sequential recommendation evaluation on Beauty, Sports, and Toys domains.}
\label{tab:sequential_eval}
\end{table*}

\begin{table*}[!htbp]
\footnotesize
\centering
\label{tab:generation_metrics}
\resizebox{1\textwidth}{!}{
\begin{tabular}{lccccc ccccc ccccc}
\toprule
\multirow{2}{*}{Methods} &
\multicolumn{5}{c}{\textbf{Beauty}} & \multicolumn{5}{c}{\textbf{Sports}} & \multicolumn{5}{c}{\textbf{Toys}} \\
\cmidrule(lr){2-6} \cmidrule(lr){7-11} \cmidrule(lr){12-16}
& BLEU2 & R-1 & R-2 & R-L & BERTS
& BLEU2 & R-1 & R-2 & R-L & BERTS
& BLEU2 & R-1 & R-2 & R-L & BERTS \\
\midrule
RR + SF (ours)&
\underline{0.473} & \textbf{15.632} & \textbf{6.298} & \textbf{12.689} & \textbf{84.831} &
\textbf{0.103} & \textbf{14.165} & \textbf{3.437} & \textbf{10.355} & \textbf{85.004} &
\textbf{0.277} & \textbf{15.558} & \textbf{4.412} & \textbf{10.765} & \textbf{85.160} \\
ChatGPT &
\textbf{1.160} & \underline{14.981} & \underline{3.041} & \underline{10.874} & \underline{82.642} &
\underline{0.023} & \underline{8.162} & \underline{1.196} & \underline{6.504} & \underline{83.410} &
\underline{0.085} & \underline{9.735} & \underline{1.433} & \underline{7.342} & \underline{83.673} \\
P5 &
0.006 & 2.162 & 0.120 & 2.070 & 8.535 &
0.001 & 2.577 & 0.113 & 2.296 & 9.984 &
0.001 & 2.407 & 0.113 & 2.176 & 8.596 \\
\bottomrule
\end{tabular}
}
\caption{Performance comparison for explanation generation across three domains. BLEU2: bi-gram precision; R-1/R-2/R-L: ROUGE scores for unigram, bigram, and longest sequence matches; BERTScore: semantic similarity.}
\label{tab:explanation generation}
\end{table*}

\subsection{Training Details}
In the \textsc{Re-Ranking} stage for candidate item selection, scores are computed using weights of $\alpha = 0.4$, $\beta = 0.4$, and $\gamma = 0.2$, and the top 30 items are extracted and fed into the LLM prompt.

\section{Experimental Results}
\subsection{Results on Recommendation Tasks}
The proposed framework was evaluated across three key recommendation tasks—direct recommendation, sequential recommendation, and explanation generation. The direct recommendation task involves predicting the Top-N items, including the ground-truth, from a pool of 100 candidates. The sequential recommendation task aims to predict the next likely item based on the user's purchase history. As shown in Table~\ref{tab:direct_eval} and Table~\ref{tab:sequential_eval}, our proposed system (RR+SF) consistently outperformed all baseline models across all domains. This demonstrates that, unlike conventional models limited to static inference or pretraining-based reasoning, our framework benefits from an active processing structure that combines \textsc{Re-Ranking} and \textsc{Self-Feedback}, resulting in more robust and adaptive performance.

\hspace*{1em}The explanation generation task was introduced to go beyond item recommendation and provide users with clear, natural language explanations for the recommendations. Specifically, the LLM generates explanations based on the relationship between the user and item profiles, focusing on relevant aspect categories. Examples of generated explanations are shown in Figure~\ref{fig:eg example}. Since this task is conditioned on the final recommendation result and the aspect profile of each item, \textsc{Re-Ranking} and \textsc{Self-Feedback} influence the outcome only indirectly. Thus, we compare the generation quality of RR+SF against existing LLM-based baselines. As shown in Table~\ref{tab:explanation generation}, the proposed model achieved the highest performance across all domains.
\subsection{Ablation Study on \textsc{Re-Ranking} and \textsc{Self-Feedback} Modules}
\begin{figure}[t]
  \begin{center}
    \includegraphics[width=\linewidth]{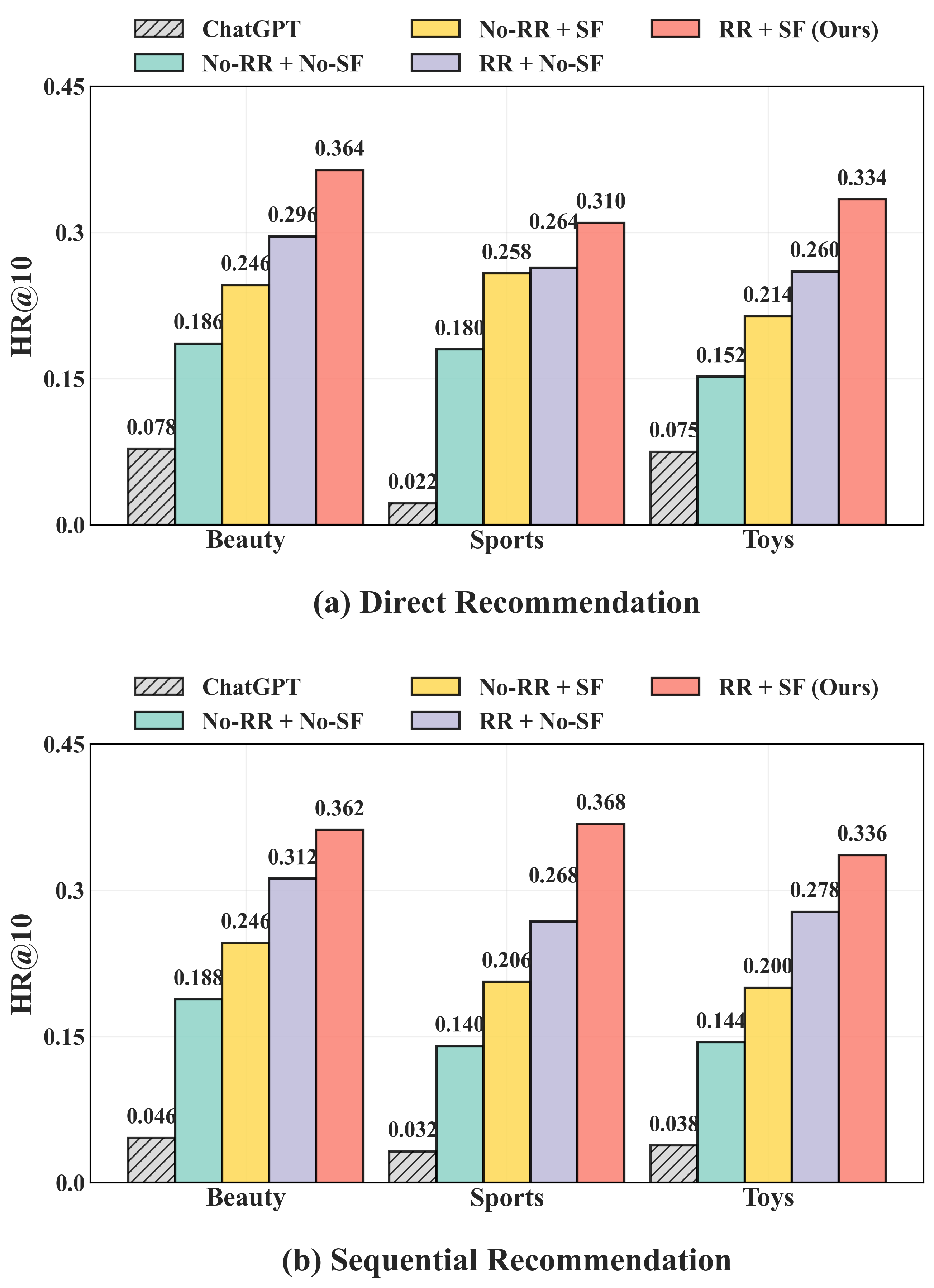}
  \end{center}
    \caption {Performance comparison (HR@10) across Beauty, Sports, and Toys domains for four model variants and \texttt{ChatGPT}. RR + SF (ours) consistently outperforms all baselines, while \texttt{ChatGPT} exhibits limited effectiveness, particularly in sequential recommendation.
}
\setlength{\abovecaptionskip}{2pt}
\setlength{\belowcaptionskip}{0pt}
  \label{fig:comparison}
\end{figure}
To quantitatively analyze the effectiveness of the two core components of our proposed system—\textsc{Re-Ranking} and \textsc{Self-Feedback}—we conducted experiments on the following four combinations. All experiments were performed under the same dataset, prompt structure, and LLM architecture. Detailed descriptions of the prompts used in each setting are provided in Appendix~\ref{sec:Prompt}. The results are summarized in Table~\ref{tab:direct_eval}, Table~\ref{tab:sequential_eval}, and Table~\ref{tab:explanation generation}.
In the No-RR+SF setting, \textsc{Re-Ranking} is omitted and recommendations are generated in the original candidate order, followed by the application of \textsc{Self-Feedback}. In RR+No-SF, only \textsc{Re-Ranking} is applied without any feedback on the recommendation outcome. The No-RR+No-SF setting disables both modules and represents the most basic recommendation structure that directly infers over unranked candidates.
Across all domains and tasks, the RR+SF configuration—where both \textsc{Re-Ranking} and \textsc{Self-Feedback} are applied—achieved the best performance. In the direct recommendation task, RR+SF showed relative improvements over No-RR+No-SF of 95.7\% in Beauty, 72.2\% in Sports, and 119.7\% in Toys. In the sequential recommendation task, the improvements were 92.6\%, 162.9\%, and 133.3\%, respectively.

\hspace*{1em}To visualize the individual and combined effects of \textsc{Re-Ranking} and \textsc{Self-Feedback}, Figure~\ref{fig:comparison} presents HR@10 scores from two perspectives. The figure compares the performance of all four configurations and \texttt{ChatGPT} across the Beauty, Sports, and Toys domains, clearly showing that RR+SF (ours) consistently outperforms all other baselines.

A notable observation is that the simple prompt-based LLM approach (\texttt{ChatGPT}) yields the lowest performance in all domains, demonstrating the superiority of leveraging aspect-based user and item profiles. In particular, the direct recommendation task in the Sports domain reveals an approximately 8× performance gap between \texttt{ChatGPT} (0.022) and No-RR+No-SF (0.180), highlighting the especially pronounced shortcomings of prompt-only methods in this task.

\hspace*{1em}The RR+No-SF and No-RR+SF configurations allow for a clear analysis of the individual contributions of each module. RR+No-SF achieved substantial improvements over No-RR+No-SF across all domains, indicating that the \textsc{Re-Ranking} module plays a more significant role in overall performance. Specifically, \textsc{Re-Ranking} sorts the candidate items based on multi-aspect profile similarity, category overlap, and popularity, selecting the top 30 most informative items as input to the LLM. This enables the model to perform more accurate reasoning within the limited context window.
Similarly, No-RR+SF also outperformed No-RR+No-SF in all domains, demonstrating the effectiveness of the \textsc{Self-Feedback} module. When recommendations are suboptimal, \textsc{Self-Feedback} automatically adjusts the scoring criteria and re-invokes inference, mimicking real user behaviors such as re-searching or re-filtering, and enabling \textit{iterative refinement}.
Finally, RR+SF achieved the largest performance gains compared to No-RR+No-SF, empirically demonstrating that the two modules work synergistically, producing a greater effect than their individual contributions alone. These results confirm that using both modules together yields the strongest performance and highlight a key structural advantage over conventional systems that rely solely on static inference.
\subsection{Human Evaluation}
\begin{table}[ht]
\centering
\footnotesize
\renewcommand{\arraystretch}{1.2}
\begin{tabular}{lcccl}
\hline
\multirow{2}{*}{Methods} & \multicolumn{3}{c}{Evaluator} & \multirow{2}{*}{Average} \\
\cline{2-4}
 & Eva\_1 & Eva\_2 & Eva\_3 & \\
\hline
P5 & 0.34 & 0.34 & 0.36 & 0.35 \\
ChatGPT & 0.00 & 0.00 & 0.00 & 0.00 \\
RR + SF (Ours) & 0.66 & 0.66 & 0.64 & \textbf{0.65} \\
\hline
\end{tabular}
\caption{Human evaluation results of explanation quality, rated by three independent evaluators. RR + SF (Ours) significantly outperforms P5 and ChatGPT in terms of average human preference.}
\label{tab:human_eval}
\end{table}
Since the linguistic quality and persuasiveness of recommendation explanations are difficult to fully evaluate using automatic metrics alone, we additionally conducted a human evaluation. Specifically, three independent evaluators (Evaluator 1, 2, and 3) were asked to compare the explanations generated by \texttt{P5}, \texttt{ChatGPT}, and our proposed model (RR+SF) across 50 test cases. Each evaluator ranked the three explanations for each case, and Table~\ref{tab:human_eval} reports the percentage of times each method was selected as the top-1 explanation by each evaluator. The results show that the proposed model was consistently rated highest by all evaluators. This indicates that our model is able to generate more specific and persuasive explanations by grounding its reasoning in aspect-level user preferences.

\section{Conclusion}
In this study, we propose \madrec, a Multi-Aspect Driven LLM Agent for explainable and personalized recommendation. The framework extracts multidimensional aspect information from user reviews in an unsupervised manner and generates structured user and item profiles that reflect diverse preference dimensions. By combining unsupervised multi-aspect learning with an LLM-based agent architecture, \madrec~identifies aspect terms and categories, summarizes category-specific content, and constructs interpretable profiles. These profiles are refined using a \textsc{Re-Ranking Tool} and provided as input to the LLM, while the \textsc{Self-Feedback} module dynamically adjusts recommendation criteria based on previous outputs, enabling iterative improvement. Evaluations on three recommendation tasks show that \madrec~consistently outperforms traditional and LLM-based baselines, not only in accuracy but also in explainability. Human evaluation further confirms that our model delivers the most persuasive explanations. In future work, we plan to improve the adaptability and interactivity of the system by incorporating user feedback-driven learning and integrating external tools.


\section{Limitations}
This study proposes an LLM-based active recommendation framework and demonstrates meaningful performance improvements across various recommendation tasks. Nevertheless, several limitations remain.
First, the multi-stage inference pipeline introduced by \textsc{Re-Ranking} and \textsc{Self-Feedback} may increase computational cost and response time, requiring further optimization for real-time applications.
Second, aspect-based inputs can be constrained by context length limits, necessitating input compression or selection strategies.
Third, while \textsc{Self-Feedback} enables iterative recommendation, it currently relies on static criteria rather than real user responses, indicating a need for future integration with interaction logs and user behavior signals.

\section{Ethics Statement}
The training process of our proposed architecture does not involve any socially sensitive or ethically inappropriate elements. Accordingly, this study raises no ethical concerns.




\bibliography{custom}

\begin{thebibliography}{34}
\providecommand{\natexlab}[1]{#1}

\bibitem[{Bao et~al.(2023)Bao, Zhang, Zhang, Wang, Feng, and He}]{10.1145/3604915.3608857}
Keqin Bao, Jizhi Zhang, Yang Zhang, Wenjie Wang, Fuli Feng, and Xiangnan He. 2023.
\newblock \href {https://doi.org/10.1145/3604915.3608857} {Tallrec: An effective and efficient tuning framework to align large language model with recommendation}.
\newblock In \emph{Proceedings of the 17th ACM Conference on Recommender Systems}, RecSys '23, page 1007–1014, New York, NY, USA. Association for Computing Machinery.

\bibitem[{Brown et~al.(2020)Brown, Mann, Ryder, Subbiah, Kaplan, Dhariwal, Neelakantan, Shyam, Sastry, Askell, Agarwal, Herbert-Voss, Krueger, Henighan, Child, Ramesh, Ziegler, Wu, Winter, Hesse, Chen, Sigler, Litwin, Gray, Chess, Clark, Berner, McCandlish, Radford, Sutskever, and Amodei}]{brown2020languagemodelsfewshotlearners}
Tom~B. Brown, Benjamin Mann, Nick Ryder, Melanie Subbiah, Jared Kaplan, Prafulla Dhariwal, Arvind Neelakantan, Pranav Shyam, Girish Sastry, Amanda Askell, Sandhini Agarwal, Ariel Herbert-Voss, Gretchen Krueger, Tom Henighan, Rewon Child, Aditya Ramesh, Daniel~M. Ziegler, Jeffrey Wu, Clemens Winter, and 12 others. 2020.
\newblock \href {https://arxiv.org/abs/2005.14165} {Language models are few-shot learners}.
\newblock \emph{Preprint}, arXiv:2005.14165.

\bibitem[{Chase(2023)}]{LangChain2023}
Harrison Chase. 2023.
\newblock \href {https://github.com/langchain-ai/langchain} {langchain}.
\newblock GitHub repository.

\bibitem[{Chen et~al.(2019)Chen, Zhang, Wang, Ma, Li, Liu, and Ma}]{10.1145/3331184.3331192}
Chong Chen, Min Zhang, Chenyang Wang, Weizhi Ma, Minming Li, Yiqun Liu, and Shaoping Ma. 2019.
\newblock \href {https://doi.org/10.1145/3331184.3331192} {An efficient adaptive transfer neural network for social-aware recommendation}.
\newblock In \emph{Proceedings of the 42nd International ACM SIGIR Conference on Research and Development in Information Retrieval}, SIGIR'19, page 225–234, New York, NY, USA. Association for Computing Machinery.

\bibitem[{Cui et~al.(2022)Cui, Ma, Zhou, Zhou, and Yang}]{cui2022m6recgenerativepretrainedlanguage}
Zeyu Cui, Jianxin Ma, Chang Zhou, Jingren Zhou, and Hongxia Yang. 2022.
\newblock \href {https://arxiv.org/abs/2205.08084} {M6-rec: Generative pretrained language models are open-ended recommender systems}.
\newblock \emph{Preprint}, arXiv:2205.08084.

\bibitem[{Gao et~al.(2023)Gao, Sheng, Xiang, Xiong, Wang, and Zhang}]{gao2023chatrecinteractiveexplainablellmsaugmented}
Yunfan Gao, Tao Sheng, Youlin Xiang, Yun Xiong, Haofen Wang, and Jiawei Zhang. 2023.
\newblock \href {https://arxiv.org/abs/2303.14524} {Chat-rec: Towards interactive and explainable llms-augmented recommender system}.
\newblock \emph{Preprint}, arXiv:2303.14524.

\bibitem[{Gazdar and Hidri(2020)}]{10.1016/j.knosys.2019.105058}
Achraf Gazdar and Lotfi Hidri. 2020.
\newblock \href {https://doi.org/10.1016/j.knosys.2019.105058} {A new similarity measure for collaborative filtering based recommender systems}.
\newblock \emph{Know.-Based Syst.}, 188(C).

\bibitem[{Geng et~al.(2022)Geng, Liu, Fu, Ge, and Zhang}]{10.1145/3523227.3546767}
Shijie Geng, Shuchang Liu, Zuohui Fu, Yingqiang Ge, and Yongfeng Zhang. 2022.
\newblock \href {https://doi.org/10.1145/3523227.3546767} {Recommendation as language processing (rlp): A unified pretrain, personalized prompt \& predict paradigm (p5)}.
\newblock In \emph{Proceedings of the 16th ACM Conference on Recommender Systems}, RecSys '22, page 299–315, New York, NY, USA. Association for Computing Machinery.

\bibitem[{Geng et~al.(2023)Geng, Liu, Fu, Ge, and Zhang}]{geng2023recommendationlanguageprocessingrlp}
Shijie Geng, Shuchang Liu, Zuohui Fu, Yingqiang Ge, and Yongfeng Zhang. 2023.
\newblock \href {https://arxiv.org/abs/2203.13366} {Recommendation as language processing (rlp): A unified pretrain, personalized prompt \& predict paradigm (p5)}.
\newblock \emph{Preprint}, arXiv:2203.13366.

\bibitem[{Hou et~al.(2024)Hou, Zhang, Lin, Lu, Xie, McAuley, and Zhao}]{10.1007/978-3-031-56060-6_24}
Yupeng Hou, Junjie Zhang, Zihan Lin, Hongyu Lu, Ruobing Xie, Julian McAuley, and Wayne~Xin Zhao. 2024.
\newblock \href {https://doi.org/10.1007/978-3-031-56060-6_24} {Large language models are zero-shot rankers for recommender systems}.
\newblock In \emph{Advances in Information Retrieval: 46th European Conference on Information Retrieval, ECIR 2024, Glasgow, UK, March 24–28, 2024, Proceedings, Part II}, page 364–381, Berlin, Heidelberg. Springer-Verlag.

\bibitem[{Kang and McAuley(2018)}]{kang2018selfattentivesequentialrecommendation}
Wang-Cheng Kang and Julian McAuley. 2018.
\newblock \href {https://arxiv.org/abs/1808.09781} {Self-attentive sequential recommendation}.
\newblock \emph{Preprint}, arXiv:1808.09781.

\bibitem[{Kang et~al.(2023)Kang, Ni, Mehta, Sathiamoorthy, Hong, Chi, and Cheng}]{kang2023llmsunderstanduserpreferences}
Wang-Cheng Kang, Jianmo Ni, Nikhil Mehta, Maheswaran Sathiamoorthy, Lichan Hong, Ed~Chi, and Derek~Zhiyuan Cheng. 2023.
\newblock \href {https://arxiv.org/abs/2305.06474} {Do llms understand user preferences? evaluating llms on user rating prediction}.
\newblock \emph{Preprint}, arXiv:2305.06474.

\bibitem[{Li et~al.(2023)Li, Wang, Li, Fu, Shen, Shang, and McAuley}]{10.1145/3580305.3599519}
Jiacheng Li, Ming Wang, Jin Li, Jinmiao Fu, Xin Shen, Jingbo Shang, and Julian McAuley. 2023.
\newblock \href {https://doi.org/10.1145/3580305.3599519} {Text is all you need: Learning language representations for sequential recommendation}.
\newblock In \emph{Proceedings of the 29th ACM SIGKDD Conference on Knowledge Discovery and Data Mining}, KDD '23, page 1258–1267, New York, NY, USA. Association for Computing Machinery.

\bibitem[{Liu et~al.(2023{\natexlab{a}})Liu, Liu, Zhou, Lv, Zhou, and Zhang}]{liu2023chatgptgoodrecommenderpreliminary}
Junling Liu, Chao Liu, Peilin Zhou, Renjie Lv, Kang Zhou, and Yan Zhang. 2023{\natexlab{a}}.
\newblock \href {https://arxiv.org/abs/2304.10149} {Is chatgpt a good recommender? a preliminary study}.
\newblock \emph{Preprint}, arXiv:2304.10149.

\bibitem[{Liu et~al.(2023{\natexlab{b}})Liu, Lin, Hewitt, Paranjape, Bevilacqua, Petroni, and Liang}]{liu2023lostmiddlelanguagemodels}
Nelson~F. Liu, Kevin Lin, John Hewitt, Ashwin Paranjape, Michele Bevilacqua, Fabio Petroni, and Percy Liang. 2023{\natexlab{b}}.
\newblock \href {https://arxiv.org/abs/2307.03172} {Lost in the middle: How language models use long contexts}.
\newblock \emph{Preprint}, arXiv:2307.03172.

\bibitem[{Mao et~al.(2021)Mao, Zhu, Wang, Dai, Dong, Xiao, and He}]{10.1145/3459637.3482297}
Kelong Mao, Jieming Zhu, Jinpeng Wang, Quanyu Dai, Zhenhua Dong, Xi~Xiao, and Xiuqiang He. 2021.
\newblock \href {https://doi.org/10.1145/3459637.3482297} {Simplex: A simple and strong baseline for collaborative filtering}.
\newblock In \emph{Proceedings of the 30th ACM International Conference on Information \& Knowledge Management}, CIKM '21, page 1243–1252, New York, NY, USA. Association for Computing Machinery.

\bibitem[{Nakajima(2023)}]{BabyAGI2023}
Yohei Nakajima. 2023.
\newblock \href {https://github.com/yoheinakajima/babyagi} {babyagi}.
\newblock GitHub repository.

\bibitem[{OpenAI et~al.(2024)OpenAI, Achiam, Adler, Agarwal, Ahmad, Akkaya, Aleman, Almeida, Altenschmidt, Altman, Anadkat, Avila, Babuschkin, Balaji, Balcom, Baltescu, Bao, Bavarian, Belgum, Bello, Berdine, Bernadett-Shapiro, Berner, Bogdonoff, Boiko, Boyd, Brakman, Brockman, Brooks, Brundage, Button, Cai, Campbell, Cann, Carey, Carlson, Carmichael, Chan, Chang, Chantzis, Chen, Chen, Chen, Chen, Chen, Chess, Cho, Chu, Chung, Cummings, Currier, Dai, Decareaux, Degry, Deutsch, Deville, Dhar, Dohan, Dowling, Dunning, Ecoffet, Eleti, Eloundou, Farhi, Fedus, Felix, Fishman, Forte, Fulford, Gao, Georges, Gibson, Goel, Gogineni, Goh, Gontijo-Lopes, Gordon, Grafstein, Gray, Greene, Gross, Gu, Guo, Hallacy, Han, Harris, He, Heaton, Heidecke, Hesse, Hickey, Hickey, Hoeschele, Houghton, Hsu, Hu, Hu, Huizinga, Jain, Jain, Jang, Jiang, Jiang, Jin, Jin, Jomoto, Jonn, Jun, Kaftan, Łukasz Kaiser, Kamali, Kanitscheider, Keskar, Khan, Kilpatrick, Kim, Kim, Kim, Kirchner, Kiros, Knight, Kokotajlo, Łukasz Kondraciuk,
  Kondrich, Konstantinidis, Kosic, Krueger, Kuo, Lampe, Lan, Lee, Leike, Leung, Levy, Li, Lim, Lin, Lin, Litwin, Lopez, Lowe, Lue, Makanju, Malfacini, Manning, Markov, Markovski, Martin, Mayer, Mayne, McGrew, McKinney, McLeavey, McMillan, McNeil, Medina, Mehta, Menick, Metz, Mishchenko, Mishkin, Monaco, Morikawa, Mossing, Mu, Murati, Murk, Mély, Nair, Nakano, Nayak, Neelakantan, Ngo, Noh, Ouyang, O'Keefe, Pachocki, Paino, Palermo, Pantuliano, Parascandolo, Parish, Parparita, Passos, Pavlov, Peng, Perelman, de~Avila Belbute~Peres, Petrov, de~Oliveira~Pinto, Michael, Pokorny, Pokrass, Pong, Powell, Power, Power, Proehl, Puri, Radford, Rae, Ramesh, Raymond, Real, Rimbach, Ross, Rotsted, Roussez, Ryder, Saltarelli, Sanders, Santurkar, Sastry, Schmidt, Schnurr, Schulman, Selsam, Sheppard, Sherbakov, Shieh, Shoker, Shyam, Sidor, Sigler, Simens, Sitkin, Slama, Sohl, Sokolowsky, Song, Staudacher, Such, Summers, Sutskever, Tang, Tezak, Thompson, Tillet, Tootoonchian, Tseng, Tuggle, Turley, Tworek, Uribe, Vallone,
  Vijayvergiya, Voss, Wainwright, Wang, Wang, Wang, Ward, Wei, Weinmann, Welihinda, Welinder, Weng, Weng, Wiethoff, Willner, Winter, Wolrich, Wong, Workman, Wu, Wu, Wu, Xiao, Xu, Yoo, Yu, Yuan, Zaremba, Zellers, Zhang, Zhang, Zhao, Zheng, Zhuang, Zhuk, and Zoph}]{openai2024gpt4technicalreport}
OpenAI, Josh Achiam, Steven Adler, Sandhini Agarwal, Lama Ahmad, Ilge Akkaya, Florencia~Leoni Aleman, Diogo Almeida, Janko Altenschmidt, Sam Altman, Shyamal Anadkat, Red Avila, Igor Babuschkin, Suchir Balaji, Valerie Balcom, Paul Baltescu, Haiming Bao, Mohammad Bavarian, Jeff Belgum, and 262 others. 2024.
\newblock \href {https://arxiv.org/abs/2303.08774} {Gpt-4 technical report}.
\newblock \emph{Preprint}, arXiv:2303.08774.

\bibitem[{Park and Kim(2025)}]{park2025scalableunsupervisedframeworkmultiaspect}
Jiin Park and Misuk Kim. 2025.
\newblock \href {https://arxiv.org/abs/2505.09286} {A scalable unsupervised framework for multi-aspect labeling of multilingual and multi-domain review data}.
\newblock \emph{Preprint}, arXiv:2505.09286.

\bibitem[{Pérez-Almaguer et~al.(2021)Pérez-Almaguer, Yera, Alzahrani, and Martínez}]{PerezAlmaguer2021Content}
Yilena Pérez-Almaguer, Raciel Yera, Ahmad~A. Alzahrani, and Luis Martínez. 2021.
\newblock \href {https://doi.org/10.1016/j.eswa.2021.115092} {Content-based group recommender systems: A general taxonomy and further improvements}.
\newblock \emph{Expert Systems with Applications}, 179:115092.
\newblock Received 22 October 2020, Revised 14 May 2021, Accepted 12 June 2021, Available online 30 June 2021.

\bibitem[{Schick et~al.(2023)Schick, Dwivedi-Yu, Dessì, Raileanu, Lomeli, Zettlemoyer, Cancedda, and Scialom}]{schick2023toolformerlanguagemodelsteach}
Timo Schick, Jane Dwivedi-Yu, Roberto Dessì, Roberta Raileanu, Maria Lomeli, Luke Zettlemoyer, Nicola Cancedda, and Thomas Scialom. 2023.
\newblock \href {https://arxiv.org/abs/2302.04761} {Toolformer: Language models can teach themselves to use tools}.
\newblock \emph{Preprint}, arXiv:2302.04761.

\bibitem[{Schulman et~al.(2022)Schulman, Zoph, Kim, Hilton, Menick, Weng, Uribe, Fedus, Metz, Pokorny et~al.}]{schulman2022chatgpt}
John Schulman, Barret Zoph, Christina Kim, Jacob Hilton, Jacob Menick, Jiayi Weng, Juan Felipe~Ceron Uribe, Liam Fedus, Luke Metz, Michael Pokorny, and 1 others. 2022.
\newblock Chatgpt: Optimizing language models for dialogue.
\newblock \url{https://openai.com/blog/chatgpt}.
\newblock OpenAI Blog.

\bibitem[{{Significant Gravitas}(2023)}]{AutoGPT2023}
{Significant Gravitas}. 2023.
\newblock \href {https://github.com/Significant-Gravitas/AutoGPT} {Auto-gpt}.
\newblock GitHub repository.

\bibitem[{Singh et~al.(2022)Singh, Sajid, Yadav, Singh, and Saini}]{9776660}
Jagendra Singh, Mohammad Sajid, Chandra~Shekhar Yadav, Shashank~Sheshar Singh, and Manthan Saini. 2022.
\newblock \href {https://doi.org/10.1109/ICOEI53556.2022.9776660} {A novel deep neural-based music recommendation method considering user and song data}.
\newblock In \emph{2022 6th International Conference on Trends in Electronics and Informatics (ICOEI)}, pages 1--7.

\bibitem[{Tang et~al.(2024)Tang, Zhang, and Dinh}]{tang-etal-2024-aspect}
An~Tang, Xiuzhen Zhang, and Minh Dinh. 2024.
\newblock \href {https://aclanthology.org/2024.findings-eacl.96/} {Aspect-based key point analysis for quantitative summarization of reviews}.
\newblock In \emph{Findings of the Association for Computational Linguistics: EACL 2024}, pages 1419--1433, St. Julian{'}s, Malta. Association for Computational Linguistics.

\bibitem[{Touvron et~al.(2023)Touvron, Lavril, Izacard, Martinet, Lachaux, Lacroix, Rozière, Goyal, Hambro, Azhar, Rodriguez, Joulin, Grave, and Lample}]{touvron2023llamaopenefficientfoundation}
Hugo Touvron, Thibaut Lavril, Gautier Izacard, Xavier Martinet, Marie-Anne Lachaux, Timothée Lacroix, Baptiste Rozière, Naman Goyal, Eric Hambro, Faisal Azhar, Aurelien Rodriguez, Armand Joulin, Edouard Grave, and Guillaume Lample. 2023.
\newblock \href {https://arxiv.org/abs/2302.13971} {Llama: Open and efficient foundation language models}.
\newblock \emph{Preprint}, arXiv:2302.13971.

\bibitem[{Tsagkias et~al.(2021)Tsagkias, King, Kallumadi, Murdock, and de~Rijke}]{10.1145/3451964.3451966}
Manos Tsagkias, Tracy~Holloway King, Surya Kallumadi, Vanessa Murdock, and Maarten de~Rijke. 2021.
\newblock \href {https://doi.org/10.1145/3451964.3451966} {Challenges and research opportunities in ecommerce search and recommendations}.
\newblock \emph{SIGIR Forum}, 54(1).

\bibitem[{Wang and Lim(2023)}]{wang2023zeroshotnextitemrecommendationusing}
Lei Wang and Ee-Peng Lim. 2023.
\newblock \href {https://arxiv.org/abs/2304.03153} {Zero-shot next-item recommendation using large pretrained language models}.
\newblock \emph{Preprint}, arXiv:2304.03153.

\bibitem[{Wei et~al.(2019)Wei, Wang, Nie, He, Hong, and Chua}]{10.1145/3343031.3351034}
Yinwei Wei, Xiang Wang, Liqiang Nie, Xiangnan He, Richang Hong, and Tat-Seng Chua. 2019.
\newblock \href {https://doi.org/10.1145/3343031.3351034} {Mmgcn: Multi-modal graph convolution network for personalized recommendation of micro-video}.
\newblock In \emph{Proceedings of the 27th ACM International Conference on Multimedia}, MM '19, page 1437–1445, New York, NY, USA. Association for Computing Machinery.

\bibitem[{Xie et~al.(2022)Xie, Zhou, and Kim}]{xie2022decoupledinformationfusionsequential}
Yueqi Xie, Peilin Zhou, and Sunghun Kim. 2022.
\newblock \href {https://arxiv.org/abs/2204.11046} {Decoupled side information fusion for sequential recommendation}.
\newblock \emph{Preprint}, arXiv:2204.11046.

\bibitem[{Yang et~al.(2023)Yang, Chen, Jiang, Cho, Huang, and Lu}]{yang2023palrpersonalizationawarellms}
Fan Yang, Zheng Chen, Ziyan Jiang, Eunah Cho, Xiaojiang Huang, and Yanbin Lu. 2023.
\newblock \href {https://arxiv.org/abs/2305.07622} {Palr: Personalization aware llms for recommendation}.
\newblock \emph{Preprint}, arXiv:2305.07622.

\bibitem[{Yao et~al.(2023)Yao, Zhao, Yu, Du, Shafran, Narasimhan, and Cao}]{yao2023reactsynergizingreasoningacting}
Shunyu Yao, Jeffrey Zhao, Dian Yu, Nan Du, Izhak Shafran, Karthik Narasimhan, and Yuan Cao. 2023.
\newblock \href {https://arxiv.org/abs/2210.03629} {React: Synergizing reasoning and acting in language models}.
\newblock \emph{Preprint}, arXiv:2210.03629.

\bibitem[{Zhang et~al.(2021)Zhang, DING, Shui, Ma, Zou, Deoras, and Wang}]{zhang2021language}
Yuhui Zhang, HAO DING, Zeren Shui, Yifei Ma, James Zou, Anoop Deoras, and Hao Wang. 2021.
\newblock \href {https://openreview.net/forum?id=hFx3fY7-m9b} {Language models as recommender systems: Evaluations and limitations}.
\newblock In \emph{I (Still) Can't Believe It's Not Better! NeurIPS 2021 Workshop}.

\bibitem[{Zhou et~al.(2020)Zhou, Wang, Zhao, Zhu, Wang, Zhang, Wang, and Wen}]{10.1145/3340531.3411954}
Kun Zhou, Hui Wang, Wayne~Xin Zhao, Yutao Zhu, Sirui Wang, Fuzheng Zhang, Zhongyuan Wang, and Ji-Rong Wen. 2020.
\newblock \href {https://doi.org/10.1145/3340531.3411954} {S3-rec: Self-supervised learning for sequential recommendation with mutual information maximization}.
\newblock In \emph{Proceedings of the 29th ACM International Conference on Information \& Knowledge Management}, CIKM '20, page 1893–1902, New York, NY, USA. Association for Computing Machinery.

\end{thebibliography}
\clearpage 
\appendix
\section{Baseline Model Details}
\label{sec:baseline_details}
The baseline models used for comparison in this study are described in detail as follows
\subsection{Direct Recommendation Model}
\begin{itemize}
    \item \textbf{ENMF (Efficient Neural Matrix Factorization)}: A matrix factorization-based model that effectively utilizes all observed data. It offers balanced performance in terms of computational efficiency and recommendation accuracy, and shows stable results even on sparse datasets.
    \item \textbf{SimpleX}: A structurally simple collaborative filtering model that incorporates a strong cosine contrastive loss, achieving performance comparable to more complex state-of-the-art models. It is particularly advantageous in terms of efficiency and interpretability.
    \item \textbf{P5 (Personalized Prompt for Personalization)}: A prompt-based framework that handles various recommendation tasks in a text-to-text format. It effectively encodes user preferences and item characteristics using natural language processing techniques, and supports generalizable performance through multi-task learning.
    \item \textbf{ChatGPT}: A few-shot recommendation approach based on a large language model, which generates recommendations using prompts without additional fine-tuning. User preferences and item attributes are processed in natural language and provided directly in the prompt.

\end{itemize}

\subsection{Sequential Recommendation Model}

\begin{itemize}
    \item \textbf{SASRec (Self-Attentive Sequential Recommendation)}: A sequential recommendation model based on the self-attention mechanism that effectively captures important signals from users’ temporal behavior patterns. It models both short- and long-term dependencies, delivering stable performance across various sequence lengths.
    \item \textbf{S\textsuperscript{3}-Rec (Self-Supervised Sequential Recommendation)}: A model that integrates multiple self-supervised learning objectives to capture rich correlations in user–item sequences. It enhances representational power by jointly optimizing item attributes, sequence patterns, and user preferences.
\end{itemize}

\hspace*{1em}These baseline models represent widely adopted approaches in current recommender systems research and were selected as comparison points to fairly evaluate the performance of the proposed \madrec~framework.

\renewcommand{\thefigure}{B.\arabic{figure}}
\setcounter{figure}{0} 
\section{Example of Explanation Generation}
\label{sec:eg}
\begin{tcolorbox}
    [title={Explanation Generation Example},
    colback = cBlue_1!10, colframe = cBlue_7, coltitle=white, fonttitle=\bfseries\footnotesize,
    center title, fontupper=\footnotesize, fontlower=\footnotesize]
    Based on the user profile, the user values products that are powerful, effective, organic, and have pleasant scents, especially in hair products, with quick and efficient usage. They also prefer affordable items with high demand and utility, and they favor products that reduce frizz, smell good, and are effective for hair and skin care.
    \tcblower
    1353 : Effective for frizz reduction, pleasant scent, high utility. 
\end{tcolorbox}
\setlength{\abovecaptionskip}{3pt}
\setlength{\belowcaptionskip}{0pt}
\label{fig:eg example}
\captionof{figure}{Example of explanation generation based on a user profile and a recommended item. The upper part shows the summarized user preferences, 
and the lower part provides the natural language explanation for why item 1353 fits the user's needs.}

\onecolumn
\renewcommand{\thetable}{C.\arabic{table}}  
\setcounter{table}{0}  
\section{Aspect Term \& Category}
\label{sec:aspect_term_and_category}
\begin{table}[H]
\centering
\renewcommand{\arraystretch}{1.3}
\setlength{\tabcolsep}{4pt}
\footnotesize
\begin{tabular}{>{\raggedright\arraybackslash}p{2.6cm} >{\raggedright\arraybackslash}p{12.4cm}}
\noalign{\hrule height 1.3pt}
\rowcolor{cBlue_1} \multicolumn{2}{c}{\textbf{\large Aspect Categories and Terms from Beauty Reviews}} \\
\midrule
\rowcolor{cBlue_7!40} 
\textbf{\color{white}Aspect Category} & \textbf{\color{white}Aspect Terms} \\ 
\midrule
\rowcolor{gray!5} \textbf{Makeup} & 
shadow, liner, concealer, eyeliner, mascara, eyeshadow, brow, blush, highlighter, primer, bronzer, foundation, palette, lipgloss, powder \\ 
\midrule
\textbf{Ingredients} & 
helianthus, annuus, kernel, vegetable, hydrogenated, bran, ester, sunflower, tocopheryl, acetate, glycine, argania, soja, tocopherol, panthenol \\ 
\midrule
\rowcolor{gray!5} \textbf{Color} & 
pink, purple, nude, bright, yellow, blue, metallic, beige, gold, shimmer, red, vibrant, coral, bronze, satin \\ 
\midrule
\textbf{Hair} & 
wavy, curly, straight, braid, strand, frizzy, ponytail, layered, heat, curl, styling, volume, rinse, shampoo, comb \\ 
\midrule
\rowcolor{gray!5} \textbf{Beauty Tools} & 
file, buffer, clipper, cutter, filing, cuticle, pedicure, scissors, drill, electric, grooming, trimming, tweezer, trimmer, manicure \\ 
\midrule
\textbf{Scent} & 
musk, sandalwood, mint, aroma, vanilla, jasmine, floral, cinnamon, citrus, lavender, coconut, honey, berry, peppermint, perfume \\ 
\midrule
\rowcolor{gray!5} \textbf{Purchase} & 
amazon, cost, expensive, bargain, budget, cheaper, online, overpriced, price, seller, buy, cheapest, pricing, purchase, repurchase \\ 
\midrule
\textbf{Usage Context} & 
evening, morning, night, daily, routine, weekend, bedtime, afternoon, overnight, weekly, daytime, frequently, outdoors, workout, wedding \\
\midrule
\rowcolor{gray!5} \textbf{Improvement} & 
aging, elasticity, reduce, inflammation, dryness, soothe, wrinkle, firmness, collagen, repair, brightening, hydrate, protect, rejuvenate, calming \\
\midrule
\textbf{Packaging} & 
zipper, case, sealed, magnetic, cardboard, pocket, compartment, pouch, box, sleeve, sturdy, envelope, clip, bag, resealable \\
\midrule
\rowcolor{gray!5} \textbf{Quantity} & 
four, ten, five, six, three, twenty, ml, oz, seven, eight, two, half, nine, ounce, dozen \\
\midrule
\textbf{Usage Method} & 
cleansing, washcloth, foam, pat, massage, toner, cleanser, exfoliating, scrub, wiping, towel, rubbing, soaking, dab, blotting \\
\midrule
\rowcolor{gray!5} \textbf{Satisfaction} & 
nice, great, wonderful, awesome, impressive, excellent, amazing, fantastic, best, perfect, comfortable, attractive, exceptional, durable, unique \\
\bottomrule
\noalign{\hrule height 1.3pt}
\end{tabular}
\caption{\textbf{Extracted Aspect Categories and Terms from Beauty Reviews.} This table presents 13 distinct aspect categories automatically identified from unlabeled Beauty reviews, along with their 15 most representative terms. These categories reveal the key dimensions consumers focus on when evaluating beauty products, ranging from makeup characteristics to scent preferences and improvement effects.}
\label{tbl:beauty_aspect_term_category}
\end{table}

\begin{table}[H]
\centering
\renewcommand{\arraystretch}{1.3}
\setlength{\tabcolsep}{4pt}
\footnotesize
\begin{tabular}{>{\raggedright\arraybackslash}p{2.6cm} >{\raggedright\arraybackslash}p{12.4cm}}
\noalign{\hrule height 1.3pt}
\rowcolor{cBlue_1} \multicolumn{2}{c}{\textbf{\large Aspect Categories and Terms from Sports Reviews}} \\
\midrule
\rowcolor{cBlue_7!40} 
\textbf{\color{white}Aspect Category} & \textbf{\color{white}Aspect Terms} \\ 
\midrule
\rowcolor{gray!5} \textbf{Functionality} & 
exceptional, usability, impressive, excellent, robust, improves, outstanding, innovative, efficient, superior, practical, versatile, durable, reliable, strong \\ 
\midrule
\textbf{Brand} & 
officially, supreme, luminox, rogue, submariner, fabulous, hydroflask, omega, elite, priceless, british, multiuse, rocksolid, branding, legendary \\ 
\midrule
\rowcolor{gray!5} \textbf{Usage Context} & 
vacation, boating, campground, canoeing, concert, festival, adventure, camping, hiking, beach, picnic, weekend, outdoors, trail, snorkeling \\ 
\midrule
\textbf{Satisfaction} & 
trust, rely, willing, honest, impressed, interested, believe, aware, expect, hoping, curious, committed, determined, satisfied, pleased \\ 
\midrule
\rowcolor{gray!5} \textbf{Technology} & 
bluetooth, wireless, wifi, gps, usb, smartphone, app, network, software, touchscreen, led, charger, sensor, rechargeable, device \\ 
\midrule
\textbf{Service} & 
vendor, contacted, request, representative, emailed, distributor, supplier, seller, dealer, merchant, manufacturer, employee, shipped, customer, returned \\ 
\midrule
\rowcolor{gray!5} \textbf{Quantity/ Measurement} & 
fifty, ten, twelve, thirty, twenty, approximate, half, couple, three, quarter, two, four, dozen, maximum, ml \\ 
\midrule
\textbf{Fit} & 
stretchy, baggy, waistband, roomy, elastic, compression, breathable, padded, expandable, cinched, comfy, spacious, supportive, snug, fitted \\
\midrule
\rowcolor{gray!5} \textbf{Ease of assembly} & 
screw, clamp, fastener, tighten, bolt, nut, insert, attach, locking, quick-release, pivot, knob, hinge, mounting, latch \\
\midrule
\textbf{Durability} & 
cracking, tearing, peeling, ripping, scraping, deform, crushed, grinding, scuff, bruised, bending, chipping, snapping, abrasion, damaged \\
\bottomrule
\noalign{\hrule height 1.3pt}
\end{tabular}
\caption{\textbf{Extracted Aspect Categories and Terms from Sports Reviews.} This table presents 10 distinct aspect categories automatically identified from unlabeled Sports and Outdoors reviews, along with their 15 most representative terms. These categories highlight the key dimensions consumers consider when evaluating sports equipment, from functionality and durability to brand reputation and usage contexts.}
\label{tbl:sports_aspect_term_category}
\end{table}

\begin{table}[H]
\centering
\renewcommand{\arraystretch}{1.3}
\setlength{\tabcolsep}{4pt}
\footnotesize
\begin{tabular}{>{\raggedright\arraybackslash}p{2.6cm} >{\raggedright\arraybackslash}p{12.4cm}}
\noalign{\hrule height 1.3pt}
\rowcolor{cBlue_1} \multicolumn{2}{c}{\textbf{\large Aspect Categories and Terms from Toys Reviews}} \\
\midrule
\rowcolor{cBlue_7!40} 
\textbf{\color{white}Aspect Category} & \textbf{\color{white}Aspect Terms} \\ 
\midrule

\rowcolor{gray!5} \textbf{Purchase} & 
amazon, walmart, retailer, seller, discount, refund, sale, coupon, shipping, return, cost, price, purchase, bargain, online \\ 
\midrule

\textbf{Character} & 
avenger, batman, bumblebee, megazord, superman, spiderman, joker, catwoman, thor, jedi, darth, hulk, yoda, deadpool, venom \\ 
\midrule

\rowcolor{gray!5} \textbf{Electronic} & 
transmitter, controller, signal, frequency, mechanism, adjustment, automatic, manual, remote, controllable, electric, battery, wireless, motorized, joystick \\ 
\midrule

\textbf{Gameplay} & 
strategy, player, opponent, mission, scoring, victory, tactic, mechanic, challenge, cooperation, turn, deck, card, phase, role \\ 
\midrule

\rowcolor{gray!5} \textbf{Food} & 
pasta, pepper, cupcake, frosting, dough, icing, sprinkles, chocolate, baking, cookie, candy, pizza, cake, muffin, chocolate \\ 
\midrule

\textbf{Movement} & 
lift, slide, rotate, tilt, flip, fold, bump, push, pull, wobble, spin, lean, climb, snap, hinge \\ 
\midrule

\rowcolor{gray!5} \textbf{Age Range} & 
three, four, five, six, seven, eight, nine, ten, eleven, twelve, thirteen, fourteen, fifteen, sixteen, eighteen \\ 
\midrule

\textbf{Educational} & 
leapreader, software, ebooks, touchscreen, tablet, app, phonics, flashcard, workbook, smartphone, digital, headphone, programming, language, instructional \\ 
\midrule

\rowcolor{gray!5} \textbf{Accessories} & 
earring, headband, ribbon, scarf, necklace, bracelet, tiara, belt, glove, hat, sunglasses, pouch, mask, hairclip, pendant \\ 
\midrule

\textbf{Safety} & 
careful, cautious, supervise, supervision, guidance, injured, danger, responsible, calm, un-supervised, help, tough, nervous, stress, patience \\ 
\midrule

\rowcolor{gray!5} \textbf{Packaging} & 
fit, aligned, snap, lock, stored, attach, glued, fasten, folded, screw, stacked, sealed, labeled, carry, wrapped \\ 
\midrule

\textbf{Animal} & 
puppy, rabbit, monkey, doggy, kitty, bunny, elephant, panda, giraffe, tiger, owl, kitten, lion, bear, dolphin \\ 
\bottomrule
\noalign{\hrule height 1.3pt}
\end{tabular}
\caption{\textbf{Extracted Aspect Categories and Representative Terms from Toys Reviews.} This table presents 12 distinct aspect categories automatically identified from unlabeled Toys and Games reviews, along with their 15 most representative terms. These categories reveal the key dimensions consumers focus on when evaluating toys and games, ranging from character-based features to educational value and safety considerations.}
\label{tbl:toys_aspect_term_category}
\end{table}
\begin{table}[H]
\centering
\renewcommand{\arraystretch}{1.3}
\setlength{\tabcolsep}{6pt}
\footnotesize
\begin{tabular}{>{\raggedright\arraybackslash}p{10.5cm} >{\raggedright\arraybackslash}p{4.5cm}}
\noalign{\hrule height 1.3pt}
\rowcolor{cBlue_1} \multicolumn{2}{c}{\textbf{\large Multi-Aspect Labeling Examples}} \\
\midrule
\rowcolor{cBlue_7!40} \multicolumn{2}{l}{\textbf{\color{white}Beauty Products}} \\
\midrule
\textbf{Review} & \textbf{Multi-Aspect Category} \\
\midrule
This is the first curling iron i ever used.. and i am not planning to purchase anything else. I had a problem with the Auto on/off button at the beginning since my hand kept on pushing it by mistake, but now that i know the proper way of holding it it doesn't bother me much. I use a heat protectant so i didn't notice any damage to my hair, on the contrary, my curls ended up being soft and shiny! & \fcolorbox{gray!30}{gray!10}{Improvement}, \fcolorbox{gray!30}{gray!10}{Hair}, \fcolorbox{gray!30}{gray!10}{Purchase} \\
\midrule
Love this stuff. It's perfect for keeping my face soft and smooth, without breaking out. I especially like to use it at night. & \fcolorbox{gray!30}{gray!10}{Usage Context} \\
\midrule
I have been using this lotion for over a month now and I really like it. I researched new lotions online and this came up as dermatologist recommended so I took a chance and ordered it. It is perfect for moisturizing before putting on make-up because it does not leave the skin oily or greasy. I have sensitive skin and it seems to be perfect for me. & \fcolorbox{gray!30}{gray!10}{Makeup}, \fcolorbox{gray!30}{gray!10}{Usage Context}, \fcolorbox{gray!30}{gray!10}{Purchase} \\
\midrule
\rowcolor{cBlue_7!40} \multicolumn{2}{l}{\textbf{\color{white}Sports and Outdoors}} \\
\midrule
They work really well you can use them in any way they even work out with pull-up bars and can attach it bench and use for reverse push-ups. & \fcolorbox{gray!30}{gray!10}{Ease of assembly} \\
\midrule
I bought 3 of these to replace the key locks on my weapons. No more having to look for the key or need to turn on the light. If you preset the combo off open, you can open this in the dark. I also like the rubberized center contacts that prevent scratching the finish. & \fcolorbox{gray!30}{gray!10}{Ease of assembly}, \fcolorbox{gray!30}{gray!10}{Durability} \\
\midrule
These are hands down the best kids goggles out there as they stay put on little faces. The large coverage area also seems to give kids more security in the water and also leaves less chances of them falling off. The material is tacky without being sticky, which is great for holding on to little kids in motion. The many colors are also nice so that each kid can have their own color. They aren't indestructible and the lens can scratch so a bit of care is a good idea, but as far as kids goggles go, this is a good investment to make. & \fcolorbox{gray!30}{gray!10}{Fit}, \fcolorbox{gray!30}{gray!10}{Durability} \\
\midrule
\rowcolor{cBlue_7!40} \multicolumn{2}{l}{\textbf{\color{white}Toys and Games}} \\
\midrule
My nephew (14) suggested this game for my son (7). It couldn't have been a better suggestion. Our son loves trains and understand math well enough to enjoy this game. It's actually fun for me, too. It's really a smarter version of Monopoly. & \fcolorbox{gray!30}{gray!10}{Age Range}, \fcolorbox{gray!30}{gray!10}{Gameplay} \\
\midrule
This Sabretooth statue, is very nice and menacing. A great pick up for the Wolverine and Sabretooth admirers out there. & \fcolorbox{gray!30}{gray!10}{Character} \\
\midrule
We are all fans of TinkerBell in my house and I was thrilled to find this for my 4 year old's Innotab 2. It has great games and creative features and is by far her favorite cartridge. The best part is that more than once I have also caught my 17 year daughter playing it as well. & \fcolorbox{gray!30}{gray!10}{Age Range}, \fcolorbox{gray!30}{gray!10}{Gameplay}, \fcolorbox{gray!30}{gray!10}{Educational} \\
\bottomrule
\noalign{\hrule height 1.3pt}
\end{tabular}
\caption{\textbf{Examples of Automatically Assigned Multi-Aspect Categories for Reviews in Beauty, Sports, and Toys Domains.}
This table presents sample reviews from the Beauty, Sports, and Toys domains, along with the automatically assigned multi-aspect category labels. These labels are generated by the \textsc{Aspect Summary Tool} prior to constructing user and item profiles.}
\label{tbl:aspect_category_review}
\end{table}

\renewcommand{\thefigure}{D.\arabic{figure}}
\setcounter{figure}{0} 
\clearpage 
\section{Additional Implementation Details}
\label{sec:Prompt}

\begin{figure*}[ht]
\begin{tcolorbox}[
    title={Aspect Summary Generation Prompt},
    colback=cBlue_1!10,
    colframe=cBlue_7,
    coltitle=white,
    fonttitle=\bfseries\footnotesize,
    center title,
    fontupper=\footnotesize
]
You are an intelligent assistant that builds personalized user profiles for a recommendation system.\\
\\
Your job is to summarize what the user values most regarding the aspect ``\{aspect\}'', based on the reviews below.\\
Only extract information that is directly related to the aspect ``\{aspect\}''.\\
Ignore general praise, irrelevant sentences, or duplicated expressions.\\
\\
Focus on capturing the user's unique preferences and patterns for this aspect.\\
Summarize the user's preference or priority into one sentence within 10 words, reflecting what kind of features the user tends to like or look for.\\
\\
Reviews: \\
\texttt{"""}\\
\{combined\_text\}\\
\texttt{"""}\\
\\
Answer format: \\
Aspect: \{aspect\} \\
Summary: <Your 10-word sentence here>
\end{tcolorbox}
\captionof{figure}{Aspect-based user profiling prompt used in the \textsc{Aspect Summary Tool}.}
\label{fig:aspect summary}
\end{figure*}

\begin{figure*}[ht]
\begin{tcolorbox}
[title={Direct Recommendation Prompt},
colback = cBlue_1!8, colframe = cBlue_7, coltitle=white,
fonttitle=\bfseries\footnotesize, center title,
fontupper=\footnotesize, fontlower=\footnotesize]
You are a smart recommendation agent.\\
\\
\lbrack User Profile\rbrack  \\
Summarize what the user values in products:  
\{user\_profile\_text\}\\
\\
\lbrack Candidate Items\rbrack  \\
You are given \{len(item\_data)\} candidate items. Each includes a category and aspect-based profile summary.  \\
\\
\{item\_blocks.strip()\}\\
\\
\lbrack Task\rbrack  \\
Based on the user profile and the information for each item, select the top-\{top\_k\} items that best match the user's preferences. For each item, consider how it matches with the user's specific aspects and preferences.\\
\\
Think \textbf{step by step} before making a final decision. Choose the top \{top\_k\} products to recommend in order of priority, from highest to lowest.
\end{tcolorbox}
\begin{tcolorbox}
[title={Sequential Recommendation Prompt},
colback = cBlue_1!10, colframe = cBlue_7, coltitle=white,
fonttitle=\bfseries\footnotesize, center title,
fontupper=\footnotesize, fontlower=\footnotesize]
You are a smart recommendation agent.\\
\\
\lbrack User Profile\rbrack \\
Summarize what the user values in products:
\{user\_profile\_text\}\\
\\
\lbrack User Purchase History\rbrack \\
The user has recently purchased these items in this exact order (oldest to newest):\{recent\_items\_text\}\\
\\
\lbrack Candidate Items\rbrack \\
You are given \{len(item\_data)\} candidate items. Each includes a category and aspect-based profile summary. \\
\\
\{item\_blocks.strip()\}\\
\\
\lbrack Task\rbrack \\
Based on both the user's profile and purchase sequence/pattern, predict the next item the user is most likely to purchase. \\
The sequential pattern and evolution of the user's preferences over time.
The user's aspect-based preferences from their profile \\
\\
Think \textbf{step by step} before making a final decision, Choose the top \{top\_k\} products to recommend in order of priority, from highest to lowest.
\end{tcolorbox}
\begin{tcolorbox}
[title={Explanation Generation Prompt},
colback = cBlue_1!10, colframe = cBlue_7, coltitle=white,
fonttitle=\bfseries\footnotesize, center title,
fontupper=\footnotesize, fontlower=\footnotesize]
You are a smart recommendation agent.\\
\\
\lbrack User Profile\rbrack \\
Summarize what the user values in products: \{user\_profile\_text\}\\
\\
\lbrack Candidate Items\rbrack \\
You are given \{len(item\_data)\} candidate items. Each includes a category and aspect-based profile summary.\\
\\
\{item\_blocks.strip()\}\\
\\
\lbrack Task\rbrack \\
Based on the user profile and the information for each item, select the top-\{top\_k\} items that best match the user's preferences and explain the recommendation reason based on aspects. For each item, consider how it matches with the user's specific aspects and preferences.\\
\\
Think \textbf{step by step} before making a final decision, Choose the top \{top\_k\} products to recommend in order of priority, from highest to lowest.\\
\\
\lbrack Example\rbrack \\
Explanation: \\
- id1: Brief explanation how this item matches user's specific aspects (15 words max)
\end{tcolorbox}
\captionof{figure}{Prompt templates used for recommendation tasks, including direct recommendation, sequential prediction, and human evaluation criteria, illustrating the input structure and task instructions for each scenario.}
\label{fig:prompt}
\end{figure*}

\begin{figure*}[ht]
\begin{tcolorbox}[
    title={\textsc{Self-Feedback} Prompt for \textsc{Re-Ranking}},
    colback=cBlue_1!10,
    colframe=cBlue_7,
    coltitle=white,
    fonttitle=\bfseries\footnotesize,
    center title,
    fontupper=\footnotesize,
    fontlower=\footnotesize
]
You are a recommendation system weight analysis expert.\\
\\
\lbrack User Profile\rbrack\\
\{user\_profile\_text\}\\
\\
\lbrack Previously Recommendation\rbrack \\
\{'\textbackslash n'.join([f"- {item['title']} ({item['category']})" for item in prev\_recommended\_info])\}\\
\\
\lbrack Current Weights\rbrack \\
- Profile similarity: 0.4\\
- Category similarity: 0.4\\
- Popularity: 0.2\\
\\
Analysis:\\
1. What are the differences between the actually selected item and recommended items?\\
2. How should weights be adjusted to rank the actual item higher?\\
\\
Propose new weights in the following format:\\
\texttt{\{}
\newline
\hspace*{1em}``profile\_similarity'': 0.X,\\
\hspace*{1em}``category\_similarity'': 0.X,\\
\hspace*{1em}``popularity'': 0.X,\\
\hspace*{1em}``reasoning'': ``Explanation for weight adjustment''\\
\texttt{\}}
\end{tcolorbox}
\begin{tcolorbox}[
    title={\textsc{Self-Feedback} Prompt For No \textsc{Re-Ranking}},
    colback=cBlue_1!10,
    colframe=cBlue_7,
    coltitle=white,
    fonttitle=\bfseries\footnotesize,
    center title,
    fontupper=\footnotesize
]
You are a recommendation system that needs to improve its strategy.\\
\\
\lbrack User Profile\rbrack \\
\{user\_profile\_text\}\\
\\
\lbrack Previous Recommendation\rbrack \\
You previously recommended these items, but the customer didn't choose any of them:\\
\{'\textbackslash n'.join([f"- {item['title']} ({item['category']})" for item in prev\_recommendations\_details])\}\\
\\
\lbrack All Candidate Items\rbrack \\
\{item\_blocks.strip()\}\\
\\
\lbrack Task\rbrack \\
Since the customer didn't choose any of your previous recommendations, you need to:\\
Reconsider your recommendation strategy\\
Think about different aspects or categories that might better match the user's preferences\\
Select \{top\_k\} different items that could better satisfy the customer's needs\\
\\
Try to recommend items from different categories or with different characteristics than before.\\
\\
Choose the top \{top\_k\} products to recommend in order of priority, from highest to lowest.
\end{tcolorbox}
\captionof{figure}{\textsc{Self-Feedback} prompt templates used in \madrec~differ in feedback format depending on whether \textsc{Re-Ranking} is applied or not.}
\label{fig:Feedback prompt}
\end{figure*}

\end{document}